\documentclass[prl,aps,twocolumn,superscriptaddress]{revtex4-1}
\usepackage{latexsym}
\usepackage{graphicx}
\usepackage{verbatim}
\usepackage{multirow}
\usepackage{amsmath}
\usepackage{mathrsfs}
\usepackage{float}

\newcommand{\be}{\begin{equation}}
\newcommand{\ee}{\end{equation}}
\newcommand{\ba}{\begin{eqnarray}}
\newcommand{\ea}{\end{eqnarray}}
\newcommand{\baa}{\begin{eqnarray*}}
\newcommand{\eaa}{\end{eqnarray*}}

\def\be{\begin{equation}}
\def\ee{\end{equation}}
\def\bea{\begin{eqnarray}}
\def\eea{\end{eqnarray}}

\def\C60{A$_x$C$_{60}$}

\def\HgCu3{HgCa$_2$Cu$_3$O$_{8+y}$}
\def\HgCu4{HgBa$_2$Ca$_3$Cu$_4$O$_{10+y}$}
\def\TlCu{Tl$_2$Ba$_2$CuO$_{6+\delta}$}
\def\TlCu3{Tl$_2$Ba$_2$Ca$_2$Cu$_3$O$_{10+y}$}
\def\TlCu4{Tl$_2$Ba$_2$Ca$_3$Cu$_4$O$_{12+y}$}

\def\BiCu3{Bi$_2$Sr$_2$Ca$_{2}$Cu$_3$O$_y$}

\def\8LSCO{La$_{1.88}$Sr$_{.12}$CuO$_4$}
\def\110LNSCO{La$_{1.5}$Nd$_{0.4}$Sr$_{0.1}$CuO$_{4}$}
\def\stage4LCO{La$_{2}$CuO$_{4+\delta}$}
\def\Y248{YBa$_2$Cu$_4$O$_8$}

\def\NbSe2{NbSe$_2$}
\def\TaSe2{TaSe$_2$}
\def\TiSe2{TiSe$_2$}

\begin{document}
\title{Probing superconductivity and pairing symmetry by coherent phonons in multiorbital
superconductors }
 \author{Chandan Setty}
\affiliation{Department of Physics and Astronomy, Purdue University, West
Lafayette, Indiana 47907, USA}
\author{Jimin Zhao}
\affiliation{Beijing National
Laboratory for Condensed Matter Physics, Institute of Physics,
Chinese Academy of Sciences, Beijing 100090,
China}
\author{Jiangping Hu}\email{jphu@iphy.ac.cn}
\affiliation{Beijing National
Laboratory for Condensed Matter Physics, Institute of Physics,
Chinese Academy of Sciences, Beijing 100090,
China}
\affiliation{Department of Physics and Astronomy, Purdue University, West
Lafayette, Indiana 47907, USA}
\affiliation{Collaborative Innovation Center of Quantum Matter, Beijing, China}
\begin{abstract}
We show that the phase information contained in coherent phonon oscillations generated by a laser pulse in a multi-
orbital superconductor
can be used as an experimental tool to probe superconductivity and pairing symmetries. The phase
difference between the normal and superconducting states is proportional to the superconducting
order parameter just below the superconducting transition temperature, $T_c$. It also exhibits different behaviors for superconducting states with
different pairing symmetries. In particular, if there is an orbital-dependent internal sign change state, the
phase difference can have a  discontinous jump below $T_c$.
\end{abstract}

\maketitle
\textit{Introduction:}   The superconductivity in a multi-band electronic system can be extremly rich and complex. Many recently discovered correlated electron systems belong to this category of multi-band superconductors. For example, iron-based superconductors discovered six years ago\cite{Hosono2008} have multiple Fermi surfaces and their bands near the  Fermi level are  attributed to all five $d-$orbitals. These materials exhibit a variety of intriguing properties associated with  all of  the degrees of freedom including charge, orbital, spin  and lattice\cite{Johnston2010}, which can, in principle, lead to  many possible  novel superconducting states\cite{Mazin2011}.   


While theoretically, a multi band structure is a fertile ground for new physics,  in experiments,  it is still very difficult  to detect them  and determine their mechanisms because of the involvement of  the multi-degrees of freedom.   Many experimental observations can be subject to multiple interpretations;  for example,   in iron-based superconductors \cite{Stewart2011}, the interplay among electronic nematicity, magnetism and orbital ordering is still a subject of active research\cite{Schmalian2014,Dagotto2012}.  The  pairing symmetry of the superconducting state, arguably the most important property, is still controversial and highly debated\cite{Mazin2011}.  While the magnitude of the superconducting order parameter can be directly probed by many experimental techniques, such as angle-resolved photoemission spectra (ARPES) and scanning tunneling microscopy(STM), there are few good direct probes to the  phase distribution of the superconducting order parameter across their multiorbital Fermi surface.  In particular,  when the phase distribution is not enforced by the symmetry of the superconducting state, as the case stands in many theoretically proposed states in iron-based superconductors, the phase sensitive junction techniques\cite{Harlingen1995} that determined the $d-$wave pairing symmetry in cuprates is also not applicable.

Since the last couple of decades, ultrafast pump-probe spectroscopy has played an increasing role in probing the superconducting ground state, with the high $T_c$ Cuprates having grabbed much of the attention \cite{Eckstein2005,Parmigiani2009,Rubhausen2009,Wolf2007,Zewail2008,Onellion2002,Kurz1992,Mihailovic2008,Ryan2005,Sugai2005,Lee2007, Mihailovic2005,Liang2013,Evetts1997,Hardy2004,Hardy2002,Hardy2003,Mihailovic1999,Mihailovic2002,Mihailovic1999-PRL}, along with a few experiments performed on multiorbital iron superconductors \cite{Ferdeghini2013, Vasiliev2012, Vasiliev2012-PRL, Taylor2010,Sood2012, Marsi2010,Marsi2009, Mihailovic2009, Nakamura2011} as well. The primary focus of most of these experiments has been the measurement of relaxation times that can be extracted from the behavior of the change in reflectivity $\frac{\Delta R}{R}$ of the probe pulse as a function of the delay time $\delta$ between the pump and the probe. From this, one can indirectly obtain information about the strength of the electron-phonon couplings and their anisotropies \cite{Wolf2007,Zewail2008}, density of photoexcited quasiparticles \cite{Mihailovic2005, Mihailovic1999}, pseudo and superconducting gaps \cite{Mihailovic1999}, and signatures of the origin of the superconducting interaction \cite{Mihailovic2008}. However, even though coherent phonon oscillations in ultrafast experiments have been generated \cite{Kurz1992, Nakamura2011}  and studied \cite{Marsi2009,Sood2012} for a while now, only a few experimental works address the role of the superconducting phase on these oscillations and no theoretical background has been laid. 

In this Letter, we  show that the phase of these coherent phonon oscillations contains useful information about the superconducting phase and its pairing symmetry; in particular, we  show that the difference in the phase of the oscillations between the normal and superconducting state is proportional to the superconducting gap, and in certain scenarios, can help distinguish the sign change of superconducting orders on different bands. Thus, the coherent phonons can act as a new experimental probe of   superconducting symmetries. 

The coherent phonon amplitude mode with wave vector $q$ is described by the driven harmonic oscillator\cite{Merlin1997,Sabbah2007}
\begin{equation}
\frac{d^2 Q_q}{dt^2} + 2 \beta \frac{dQ_q}{dt} + \Omega^2 Q_q =  F(t)
 \end{equation}
where $Q_q$ is the amplitude of the phonon mode, $\Omega$ is the frequency of the oscillator, $\beta$ is the damping parameter and $F(t)$ is the driving force. The solution to the above equation is given by
\begin{equation}
Q_q(t) = \mathcal{A} e^{-\beta t} cos(\tilde{\Omega} t + \Gamma_{ph})
\end{equation}
where $\tilde{\Omega} = \sqrt{\Omega^2 - \beta^2}$ and $\mathcal{A}$ is the amplitude of the oscillation which is proportional  to the magnitude of the driving force $F$. For simplicity, we will ignore any effect of damping. In such a case, the phase of the phonon oscillation $\Gamma_{ph}$ is given by \cite{Sabbah2007}
\begin{equation}
Tan(\Gamma_{ph}) = \frac{Im(i F(-\Omega))}{Re(i F(-\Omega))}.
\end{equation}

The driving force $F(t)$ can be derived microscopically under reasonable approximations.  Consider a general  Hamiltonian that describes the  physical processes  in an ultrafast pump-probe experiment given by
\begin{equation}
H = H_e + H_p + H_{e-p} + H_{e-l}(t),
\end{equation} 
where $H_e$, $H_p$, $H_{ep}$ and $H_{el}(t)$ are electronic, phononic, electron-phonon coupling and electron-pulse interaction parts respectively\cite{Merlin1997}.  In a superconducting state,   the electronic part, $H_e$, is given by the general BCS form
\begin{eqnarray}
H_e &=& \sum_{k\sigma\alpha\beta} \epsilon_{k\sigma\alpha\beta} c_{k\sigma\alpha}^{\dagger}c_{k\sigma\beta} + \sum_{k\alpha} \Delta_{k\alpha} c_{k\alpha\uparrow}^{\dagger}c_{-k\alpha\downarrow}^{\dagger}
\end{eqnarray}
where $\alpha $ and $\sigma$ are the orbital and spin index.  We take the standard form for $H_p = \frac{1}{2}\sum_{q} (P_q^2 + \Omega_q^2 Q_q^2) $ and $
H_{ep} = \sum_{kq\alpha\alpha'} \xi_{\alpha\alpha'} Q_q c_{k\alpha}^{\dagger}c_{k+q\alpha'}+ h.c$ where $Q_q$ and $P_q$ are the canonical coordinates and momenta.  The time dependent electron-laser pulse interaction is given by $H_{el}(t)= \sum_{kq\alpha\alpha'}V_{\alpha\alpha'}(t)  c_{k\alpha}^{\dagger}c_{k+q\alpha'}+ h.c. $ with  $V_{\alpha \alpha'}(t) = \frac{e}{m}\int dr_i \phi_{\alpha}(r_i)^* [\vec{A}(t)\cdot\vec p_i] \phi_{\alpha'}(r_i) $.  As the coherent phonons are generated at $q=0$ and the momentum of the light is much smaller than  the electron momentum,  we can set $q=0$ in all above Hamiltonians.  

 We consider  parameters in a typical femtosecond pump-probe experiment. The pump pulse (central frequency $\omega_o \sim 375 THz$) has a width of $\tau \sim 80 fs$ and a relatively broad spectral width of the order of $\Delta \nu\sim 5-10 THz$. Such a spectral width is just enough to excite the lowest energy optical phonon mode whose energy is around $\Omega\sim 5 THz$. To ensure that the phonon oscillations are properly resolved in time, the width of the pump laser pulse satisfies the condition $\tau<< \Omega^{-1}$.  

 The average force driving the coherent phonon oscillations is given by $F(t) = - \partial \langle H_{e-p} \rangle(t)/ \partial Q_{\vec q}$.
\begin{figure}[h!]
\caption{\label{Scattering}A cartoon plot showing the toy band structure used to illustrate the scattering between superconducting bands close to the Fermi level. A quasiparticle is light scattered (solid wavy line) from an occupied band state $a$ to an empty state in band state $b$ and light scattered again from band state $b$ to another band state $c$. Finally the quasiparticle makes a transition back to the band state $a$ by scattering with a phonon (dashed-dotted line). The energy scale on the vertical axis is of the order of the superconducting gap. }
\includegraphics[width=0.45\textwidth]{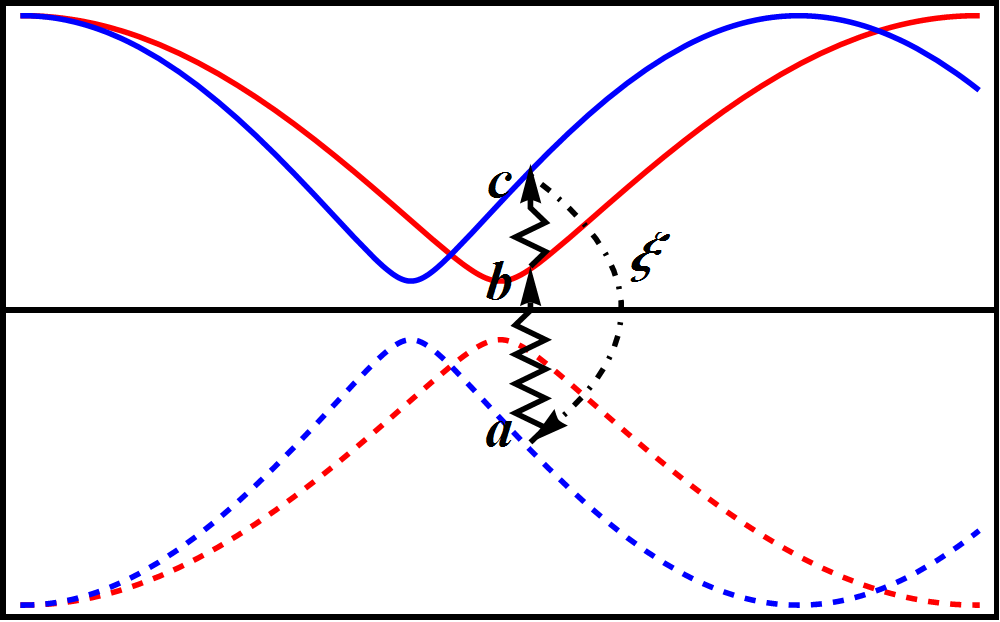}
\end{figure} 
Here, $\langle ...\rangle$ denotes an ensemble average over eigen states of $H - H_{e-l}(t)$ time evolving in $H_{e-l}(t)$ perturbatively. 
In lines with the authors in ref \cite{Sabbah2007}, we assume that the electric field is spatially homogeneous and a gaussian centered around $\omega_o$. Thus the electric field product $E(\omega)E(\omega+\Omega)$ is strongly peaked at $\omega_o - \Omega/2$. This leads to an expression for the driving force\cite{Merlin1997, Sabbah2007}
\begin{eqnarray}\nonumber
F(\Omega) &=& \frac{-C}{\left(\omega_o - \frac{\Omega^2}{4}\right)}\sum_{\vec k m n} \Biggl(\frac{\tilde{\xi}_{In}\tilde{V}_{nm}(\vec k)\tilde{V}_{mI}(\vec k)}{(\omega_{nI} - \Omega - i g)(\omega_{mI} - \omega_o - i g)} \\  \label{DrivingForce}
&&+ \frac{\tilde{\xi}_{nI}\tilde{V}_{mn}(\vec k) \tilde{V}_{Im}(\vec k)}{(\omega_{nI} + \Omega + i g)(\omega_{mI} - \omega_o + i g)} \Biggr).
\end{eqnarray}
Here $n,m,I$ are band states, $C$ is an unimportant constant, $g$ contributes to the optical absorption, and $\vec k$ is the crystal momentum. We have defined $\omega_{nI} \equiv \omega_{nI}(\vec k) = \omega_n - \omega_I$, where $\omega_n$ is the energy of band $n$ with momentum $\vec k$. The tilde sign above the matrix elements denotes the respective quantities written in the band basis. In the expression for $F(\Omega)$, we have assumed that the laser frequency is the largest energy scale in the problem. Therefore, we have chosen to keep the most resonant terms by ignoring a third term which has a denominator proportional to $\omega_o^2$.\\
\newline
\textit{a) Two band case - analytic result}: Our goal in this section is to study the phase of the coherent phonon oscillations ($\Gamma_{ph}$) across $T_c$ for a generic two orbital model. Our model comprises intraorbital hoppings $\epsilon_{1}(\vec k), \epsilon_{2}(\vec k)$, and the interorbital hopping $\epsilon_{12}(\vec k)\equiv m_k $. For analytic simplicity, we choose the two orbitals to have the same complex gap order parameter $|\Delta| e^{i \phi}$. This condition will be relaxed in the next section where we apply numerics. For the electron phonon couplings, we only keep non-zero matrix elements 
 for the coupling between the two different orbitals ($\xi'$) and coupling between superconducting particle-hole bands ($ \xi e^{i \phi}$).
As the electronic response to the laser pulse is a very fast process,  we can assume that the phonon is not activated during the laser pulse excitation. Such an approximation is easy to justify considering that the fast moving electrons have a larger effect on the slow moving nuclei than the other way around$-$ an analogue of the Born-Oppenheimer approximation in atomic physics.  In this case,  the lowest order effect of the electron-phonon coupling is in the form of the driving force,  $F(t) = - \partial \langle H_{e-p} \rangle(t)/ \partial Q_{\vec q}$. 
\newline
\newline
We now  proceed with our calculation of the average driving force $F(t)$. To bring out the physics essential for our discussion, we consider a  scattering process illustrated in the cartoon in Fig \ref{Scattering}. A quasiparticle in  the state $a$ is scattered by a photon to the  empty state $b$ above the Fermi level, and then scattered again into another empty state $c$ by a second photon. Finally, the quasiparticle is scattered back to its original state $a$ through a phonon or a series of phonons.
 We  explicitely evaluate the matrix element product $\tilde{\xi}_{ac}\tilde{V}_{cb}(\vec k)\tilde{V}_{ba}(\vec k)$ for such a process so that other similar scattering processes can be determined analogously. To do this, we first have to perform an unitary transform into the orbital basis and then use the formulas described in \cite{Pedersen2001} for tight binding matrix elements. We can write the above matrix element product as
\begin{eqnarray}
\tilde{\xi}_{ac}\tilde{V}_{cb}(\vec k)\tilde{V}_{ba}(\vec k) &=&  \Delta\frac{f(\theta_k)}{x_+^2 x_-^4} (\mathscr{E_+} + \mathscr{E_-})(\mathscr{E_+} \mathscr{E_-} - \Delta^2)\\
&& \times \left[\xi (\mathscr{E_-}^2 - \Delta^2) + \xi' sin 2\theta_k (2 \Delta \mathscr{E_-})\right],\nonumber
\end{eqnarray}
where $f(\theta_k) = -(\partial m_k)^2 cos^2 2 \theta_k $, $x_{\pm} = \sqrt{\Delta^2 + \mathscr{E}_{\pm}^2}$, $\mathscr{E}_\pm = \epsilon_{\pm}(\vec k) + E_{\pm}(\vec k)$, 
with the band angle $tan 2\theta_k = 2 m_k/(\epsilon_{1k} - \epsilon_{2k})$, $\epsilon_{\pm}(\vec k)$ the band energies, and $E_{\pm}(\vec k) = \sqrt{\Delta^2 + \epsilon_{\pm}(\vec k)^2}$.
From the above expression for the matrix element product, we can separate the most dominant contributions from different regions of the Brillouin zone. We consider three different cases: (1) contributions from momentum space points far away from the fermi surface where $\epsilon_{\pm}(\vec k) >>  |\Delta|> 0$, (2) on the Fermi surface $\epsilon_+(\vec k) = 0<|\Delta|<< \epsilon_-(k)$ and finally, (3) on the Fermi surface $\epsilon_-(\vec k) = 0<|\Delta|<< \epsilon_+(k)$. We find that
\begin{eqnarray}\label{Result}
\tilde{\xi}_{ac}\tilde{V}_{cb}(\vec k)\tilde{V}_{ba}(\vec k) &=&\\
  &&\!\!\!\!\!\!\!\!\!\!\!\!\!\!\!\!\!\!\!\!\!\!\!\!\!\!\!\!\!\!\!\!\!\!\!\!\!\!\!\!\!\!\!\!\!\!\!\!f(\theta_k)\times\begin{cases}
  \xi \frac{|\Delta|}{\tilde{\epsilon}} \left( 1+ \frac{2 |\Delta|}{\mathscr{E_-}}\frac{\xi'}{\xi} s_ {2\theta}\right)&  \text {$\epsilon_{\pm} >>  |\Delta|> 0$} \nonumber\\
 \frac{ \xi }{2}  \left( 1+ \frac{2 |\Delta|}{\mathscr{E_-}}\frac{\xi'}{\xi} s_{2\theta}\right)& \text{$\epsilon_->> |\Delta|>0=\epsilon_+$}\nonumber\\
 \frac{\xi'}{2} s_{2\theta}& \text{$\epsilon_+>> |\Delta|>0=\epsilon_-$},
  \end{cases}
\end{eqnarray}
where we have defined the effective band energy $\tilde{\epsilon} = \frac{\mathscr{E_+} \mathscr{E_-}}{\mathscr{E_+}+ \mathscr{E_-}}$  and $s_{2\theta} \equiv sin 2\theta_k$. Similar expressions can be obtained for the other scattering processes.  The energy denominators appearing in the expression for the driving force in eq \ref{DrivingForce} depend quadratically on the energy gap. From this, along with the expression for the matrix element product (written in eq.\ref{Result}), we arrive at the central result of this section $-$ \textit{the coherent phonon phase encodes the behavior of the superconducting order parameter.} For small $\Delta$, the phase can be written very generally as $\Gamma_{ph} = \alpha_1 + \alpha_2 \Delta(T)$, where $\alpha_1$ and $ \alpha_2$ are constants independent of temperature. As a result, the phase difference between the superconducting and normal state is proportional to the pairing gap. We also additionally conclude that the contribution to the average driving force from the momentum points far away from the Fermi surface is of $O(\Delta/\tilde{\epsilon})$ smaller than the contribution from those close to the Fermi surfaces. However, all the regions in the Brillouin zone contribute to the phase of the oscillation to the same order. This naturally implies that for a significant driving force to be generated, we would require the frequency of the phonon mode excited ($\sim 5 -10 THz$) to be of the order of the superconducting gap. This is a condition that is hard to attain in classic  BCS superconductors, but is comfortably satisfied by high $T_c$ Cuprates and iron based superconductors. 

\begin{figure}[h!]
\caption{\label{PhasevsT}Plot showing the variation of the phase($\Gamma_{ph}$) difference between superconducting(S) and normal(N) state as a function of temperature across  $T_c$. (Left) Phase as a function of magnitude of a constant $s-$wave gap on all the three bands. (Right) Phase for different pairing forms of the gap, all the same on the three bands. The values of the electron phonon coupling is chosen as $\xi' = 0.4 eV$ for interobital, $\xi'/4$ for $xz/yz$ and $\xi'/2$ for $xy$ intraorbital coupling and the damping coefficient is chosen as $g = 0.3 eV$. The laser and phonon frequencies are fixed at $2eV$ and $0.2 eV$ respectively. }
\includegraphics[width=0.5\textwidth]{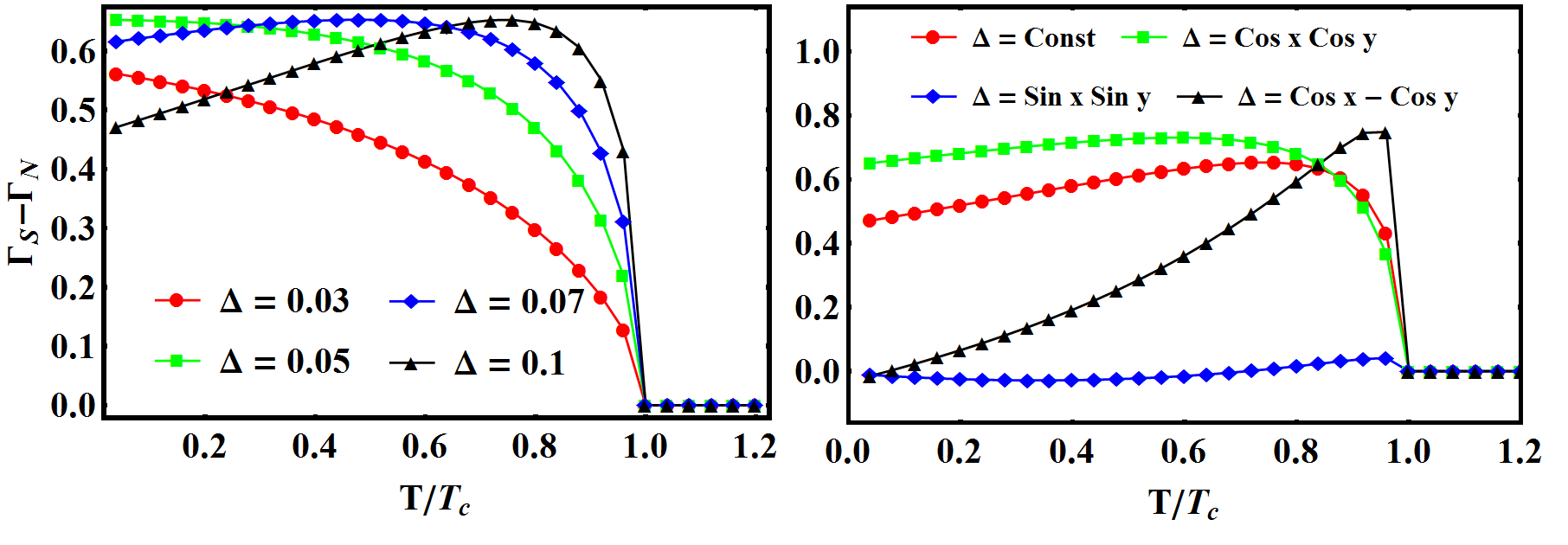}
\end{figure}

\textit{b) Three band case}:  To further test the above results, we consider  a  more realistic  band model that describes  iron based superconductors and  study the pairing symmetry dependence. We also examine any signatures that can capture the inter-orbital sign change contained in the phase of coherent phonon oscillation.  To illustrate our numerical results and maintain analytical tractability, we choose the three band model proposed by Daghofer et.al \cite{Daghofer2010}. Fig \ref{PhasevsT} shows our result for the temperature dependence plot of the phase difference $\Gamma_{S} - \Gamma_{N}$ between the superconducting and normal states across $T_c$. The phase is a constant above $T_c$ and varies below it due to the development of a superconducting gap on the Fermi surfaces.
\begin{figure}[h!]
\caption{\label{FvsgvsTWithPhaseJump} Plot showing the variation of the real part of the driving force $i F$ as a function of temperature (vertical axis) and the damping constant $g$ (horizontal axis) in the three orbital model of ref \cite{Daghofer2010}. Top row  Left (Right): case where the signs of the gap on the $xz,yz$ orbitals is the same (opposite) as that on the $xy$ orbital. The color scale represents the real part of the driving force $i F$). Center row (left and right): Cuts along different chosen values of $g$ for the corresponding color plots above them. Bottom row: The corresponding phases as a function of temperature for the $g=0.2$ case. The values of the electron phonon couplings, laser and phonon frequencies are chosen same as in fig \ref{PhasevsT}.}
\includegraphics[width=0.5\textwidth]{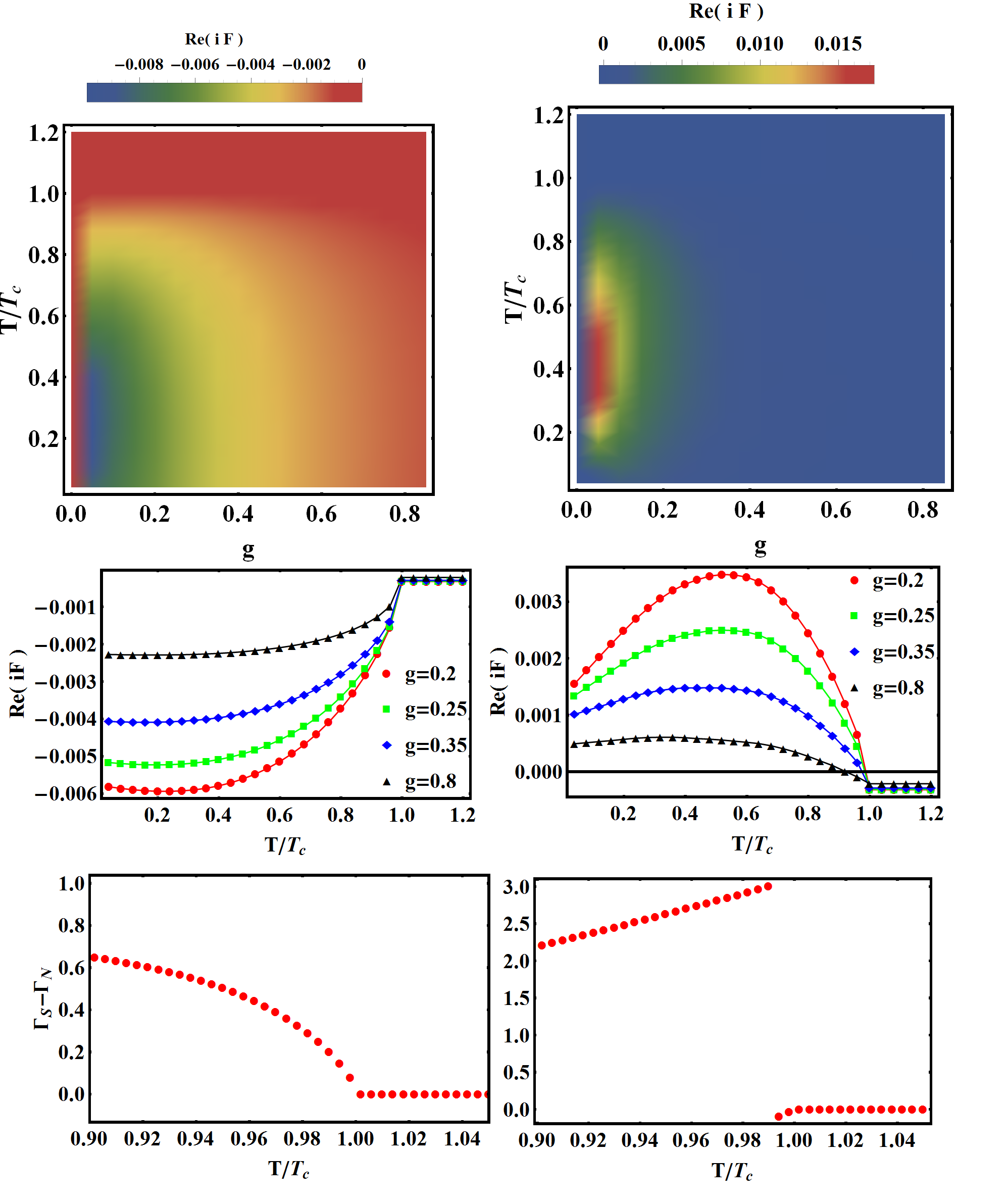}
\end{figure} 
 For a simple constant $s-$ wave pairing (Fig \ref{PhasevsT} (Left)), the variation of the phase in the SC state is maximum at $T=0$ (for small values of the gap) and follows a $linear$ dependence on $\Delta$, as was analytically derived in the previous section. However, on increasing the magnitude of $\Delta$, the change in phase develops a maximum at a temperature $0<T<T_c$ and then falls off at $T=0$ due to higher order contributions of $\Delta$.  Fig. \ref{PhasevsT} (Right) shows the plot of the phase of the oscillation as a function of temperature for different pairing symmetries. For the $s-$ wave cases, there is a substantial change in the phase between $T=0$ and $T=T_c$, whereas for the $d-$ wave cases there is little phase change between $T=0$ and $T=T_c$. In the $d-$ wave scenario, the phase sharply plunges on entering into the superconducting state. \\
\newline
Fig \ref{FvsgvsTWithPhaseJump} (top row) shows a color plot of the real part of $iF$ as a function of temperature and the absorption coefficient $g$. The left column corresponds to the case where the sign of the gap on all the three orbitals is the same (+++ case), while that on the right has a gap on the $xy$ orbital opposite in sign to that of the $xz$ and $yz$ orbitals (++- case). Fig \ref{FvsgvsTWithPhaseJump} (center row) shows cuts corresponding to different values of $g$ for both these cases. Clearly, below $T_c$, the slope of the real part of $iF$ has an opposite sign for the (+++) and (++-) cases. More specific to the three orbital model, the real part of $iF$ goes through a zero for the (++-) case and, therefore, has a $\pi$ discontinuity in the phase. On the other hand, in the (+++) scenario, the real part of $iF$ does not change sign and results in a smooth variation of phase with temperature (see fig \ref{FvsgvsTWithPhaseJump} (bottom row)).\\
\newline
To get the physics governing the numerics above, we consider the three band model with a definite sign of the gap on the $xz$ and $yz$ orbitals (denoted by $\Delta_1 = \Delta$ and $\Delta_2 = \Delta$)  and an arbitrary gap $\Delta_3$ on the $xy$ orbital. We find that for small values of $\Delta_3$, the driving force on the phonons can be written as $F(T) = \sum_{\vec{k} } \left(\tilde{\alpha}_1 (k) + sgn(\Delta \Delta_3)  \tilde{\beta}_1 (k) |\Delta_3 (T)|\right) +  i \left(\tilde{\alpha}_2 (k) + sgn(\Delta \Delta_3) \tilde{\beta}_2 (k) |\Delta_3 (T)|\right)$.
Here, $\tilde{\alpha}_i$ and $\tilde{\beta_i}$ are band structure dependent functions which can be determined for a given model. For the above model, we find that $\sum_{\vec k} \tilde{\alpha}_2(\vec k)$  and $\sum_{\vec k} \tilde{\beta}_2(\vec k)$ are both negative. This implies that when all the three orbitals have the same sign of the gap, the real part of $iF(T)$ is negative. On the other hand, if the sign change exists among  the third orbitals, the denominator becomes zero for a critical temperature and results in an observable $\pi$ phase jump.

The above results can be applied to investigate the pairing symmetries in multi-orbital superconductors.  Here we specifically discuss its applications to iron-based superconductors.  Different pairing symmetries, including s-wave\cite{Mazin2008,Kuroki2008,Chubukov2008,Hu2008,Hu2011,Onari2010} and d-wave pairing symmetries\cite{Bernevig2011-dwave, Scalapino2011}, were proposed for  different families of iron-based superconductors. Even within the s-wave pairing symmetry, there are a variety of possibilities on the internal sign changes, including the sign changes between different pockets (so called $s^\pm$\cite{Mazin2008,Kuroki2008,Chubukov2008,Hu2008}) and between bands featured by different orbitals(so called orbital-dependent $S^\pm$ or antiphase-$s^\pm$\cite{Hu2012-Xiaoli,Hu2014,Hu2013-OddParity,Kotliar2014}). Our results suggest that  the phase of coherent phonons should have distinct behaviors with respect to the  $s\pm$, antiphase-$s^\pm$ and d-wave states. In particular, as shown in fig \ref{FvsgvsTWithPhaseJump},  if a phase jump can be observed below $T_c$,  it should be a smoking-gun signature for the antiphase-$s^{\pm}$ state. 

\textit{Conclusions}:  We have shown that coherent phonon oscillations can  be an experimental probe of the superconducting state and its pairing symmetry. The phase of the coherent phonon carries intrinsic information of superconducting order parameters and can be applied to determine the pairing symmetries in complex multi-orbital superconductors. 
\newline
\newline
JPH acknowledges support from grants: MOST of China (2012CB821400,2015CB921300), NSFC(11190020,91221303,11334012) and  ``Strategic Priority Research Program (B)" of the Chinese Academy of Sciences( XDB07020200). JMZ is supported by NSFC (11274372) and MOST of China (2012CB821402).
\bibliographystyle{apsrev4-1}
\bibliography{CoherentPhonons-Short}
\end{document}